\documentclass[a4paper,fleqn,usenatbib]{mnras}

\usepackage{newtxtext,newtxmath}

\usepackage[T1]{fontenc}
\usepackage{ae,aecompl}

\usepackage{graphicx}	
\usepackage{amsmath}	
\usepackage{amssymb}	
\usepackage{graphicx}	
\usepackage{amsmath}	
\usepackage{amssymb}	
\usepackage{lscape}
\usepackage[utf8]{inputenc}
\usepackage{siunitx}
\usepackage[english]{babel}
\usepackage[T1]{fontenc}
\usepackage{ae,aecompl}
\usepackage{textcomp}
\usepackage{soul}

\title[Blazar classification with neural network]{Classification of Blazar Candidates of Uncertain Type from the  Fermi LAT 8-Year Source Catalog with an Artificial Neural Network}

\author[M. Kova{\v{c}}evi{\'{c}} et al.]
{
M. Kova{\v{c}}evi{\'{c}}$^{1}$\thanks{E-mail: milosh.kovacevic@gmail.com},
G. Chiaro$^{2}$\thanks{E-mail: graziano.chiaro@inaf.it},
S. Cutini$^{1}$,
G. Tosti$^{3}$
\\
\\
$^{1}$INFN -- Istituto Nazionale di Fisica Nucleare Sez. Perugia, I-06123 Perugia, Italy\\
$^{2}$INAF -- Istituto di Astrofisica Spaziale e Fisica Cosmica, I-20133 Milano, Italy\\
$^{3}$Dipartimento di Fisica e Geologia, Univ. degli Studi di Perugia, I-06123 Perugia, Italy\\
}

\date{Accepted 2020 February 3. Received 2020 February 2; in original form 2020 January 7.}

\pubyear{2020}

\begin{document}
\label{firstpage}
\pagerange{\pageref{firstpage}--\pageref{lastpage}}
\maketitle

\begin{abstract}  
The \textit{Fermi} Large Area Telescope (LAT) has detected more than 5000 $\gamma$-ray sources in its first 8 years of operation. More than 3000 of them are blazars. About 60\% of the \textit{Fermi}-LAT blazars are classified as BL Lacertae objects (BL Lacs) or Flat Spectrum Radio Quasars (FSRQs), while the rest remain of uncertain type. The goal of this study was to classify those blazars of uncertain type, using a supervised machine learning method based on an artificial neural network, by comparing their properties to those of known $\gamma$-ray sources. Probabilities for each of 1329 uncertain blazars to be a BL Lac or FSRQ are obtained. Using 90\% precision metric, 801 can be classified as BL Lacs and 406 as FSRQs while 122 still remain unclassified. This approach is of interest because it gives a fast preliminary classification of uncertain blazars. We also explored how different selections of training and testing samples affect the classification and discuss the meaning of network outputs.
\end{abstract}


\section{Introduction}

Blazars are active galactic nuclei (AGNs) with a radio-loud behaviour and a relativistic jet pointing towards the observer \citep{Abdo01, fra}. These sources are divided into two main classes: BL Lacertae objects (BL Lacs) and Flat Spectrum Radio Quasars (FSRQs), which show very different optical spectra. FSRQs have strong, broad emission lines,  while BL Lacs show mostly weak or no emission lines. Compact radio cores, flat radio spectra, high brightness temperatures, superluminal motion, high polarization, and strong and rapid variability are also commonly found in BL Lacs and FSRQs. Blazars emit variable, non-thermal radiation across the whole electromagnetic spectrum, featuring components forming two broad humps in a $\nu f{_\nu}$ representation, where $\nu$ is the observing frequency and  $f{_\nu}$ the spectral energy density. The low-energy hump is attributed to synchrotron radiation, and the high-energy one is usually thought to be due to inverse Compton radiation \citep{Ghisellini}. 

The \textit{Fermi} Large Area Telescope (LAT) has been continuously observing the $\gamma$-ray sky since 2008 August in the 100 MeV--300 GeV energy range. The latest \textit{Fermi}-LAT catalog is the LAT 8-year Source Catalog \textit{4FGL} \citep{4fgl}, which lists 5066 $\gamma$-ray sources, about 2000 more than the previous 3FGL catalog \citep{3fgl}, which was based on four years of data. Out of the 5066 4FGL sources, 3131 are blazars: 1116 BL Lacs, 686 FSRQs, and 1329 blazar candidates of uncertain type (BCUs). If we compare the 4FGL with previous LAT catalogs we can see the significant increase of the number of unclassified sources. The percentage of BCUs increased from 14$\%$ in 1FGL \citep{1fgl} to 42$\%$ in 4FGL. In Table~\ref{<fgl>} we show the growth of the number of blazar sources detected by \textit{Fermi}-LAT.  The increased difficulty to have sufficiently extensive optical observation campaigns for rigorous classification of BCUs emphasizes the importance of finding alternative ways to classify blazars.

\begin{table}
\label{<fgl>}
\begin{footnotesize}
\begin{tabular}{lccccr}
\hline
\hline
\bf{Class} &\bf{1FGL} & \bf{2FGL} & \bf{3FGL}& \bf{4FGL} \\
\hline
BL Lac & 295 (44\%) & 436 (41\%) & 660 (38\%) & 1116 (36\%)\\ 
FSRQ & 278 (42\%) & 370 (35\%) & 484 (28\%) & 686 (22\%) \\
BCU & 92 (14\%) & 257 (24\%) & 573 (34\%) & 1329 (42\%)\\
\hline
Total & 665 & 1063 & 1717 & 3131\\ 
\hline
\end{tabular}
\end{footnotesize}
\caption{Blazar class distribution in \textit{Fermi}-LAT catalogs.}
\end{table} 

Since more than 1300 $\gamma$-ray sources in the 4FGL remain unassociated with any plausible source class, the full nature of almost half  the sources in the 4FGL catalog remains undetermined. Classifying BCUs remains a strategic goal not only to enlarge the number of detected BL Lacs and FSRQs but also to confirm the extragalactic background light absorption of high energy photons that will be strategic in the next Cherenkov Telescope Array (CTA) extragalactic survey, which will  investigate the physics of high-energy emission from relativistic AGN jets. For this reason, studies and methods for hunting and characterizing BCUs are very useful for the scientific community. When optical spectra or multiwavelength information needed for a rigorous classification are not available, a statistical approach to the problem, including machine learning, can be very useful for classification of BCUs. 

Machine learning is a method of recognizing patterns within data in order to achieve goals such as classification. In a type of machine learning called \textit{supervised} machine learning, an algorithm classifies unknown objects by comparing their characteristics with characteristics of known objects. 

Machine learning has been applied by \citet{ack2012, lee2012, hassan, doert2014, bflap, einecke, mirabal2016, pablo, yi, lefau, zoo, kang, kovacevic3fgl, kaur} and other studies in order to classify unassociated sources and/or BCUs from the LAT catalogs. Some of the most commonly used machine learning techniques in the above cited works, and astrophysics in general, include: Random Forest \citep{rf}, Artificial Neural Network (ANN) \citep{ann}, Support Vector Machines \citep{cor, vap}, and Boosted Decision Trees \citep{fri}.

Following \citet{bflap, zoo, kovacevic3fgl} (hereinafter \textit{C16}, \textit{S17}, \textit{K19}) in which ANN was used to classify BCUs and BCU candidates from 3FGL catalog, here we used ANN  in order to classify BCUs from the 4FGL catalog. For input parameters to the network we used $\gamma$-ray parameters present in the the 4FGL catalog\footnote{\url{https://fermi.gsfc.nasa.gov/ssc/data/access/lat/8yr_catalog/}.} which is publicly available. For ANN we used \textit{TensorFlow}\footnote{\url{https://www.tensorflow.org}. TensorFlow is an open source library for machine learning. It is relatively fast, easy to use, and transparent.} \citep{tf} which was implemented in \textit{Python}\footnote{\url{https://www.python.org/}}.

The paper is organized as follows: in Section ~\ref{<2>} we present the ANN method used. In Section ~\ref{<3>} we discuss the network outputs and caveats. In Section ~\ref{<4>} and Section ~\ref{<5>} we present and validate the results.

\section{The ANN method}
\label{<2>}

\begin{figure}
\begin{center}
\includegraphics[width=.35\textwidth]{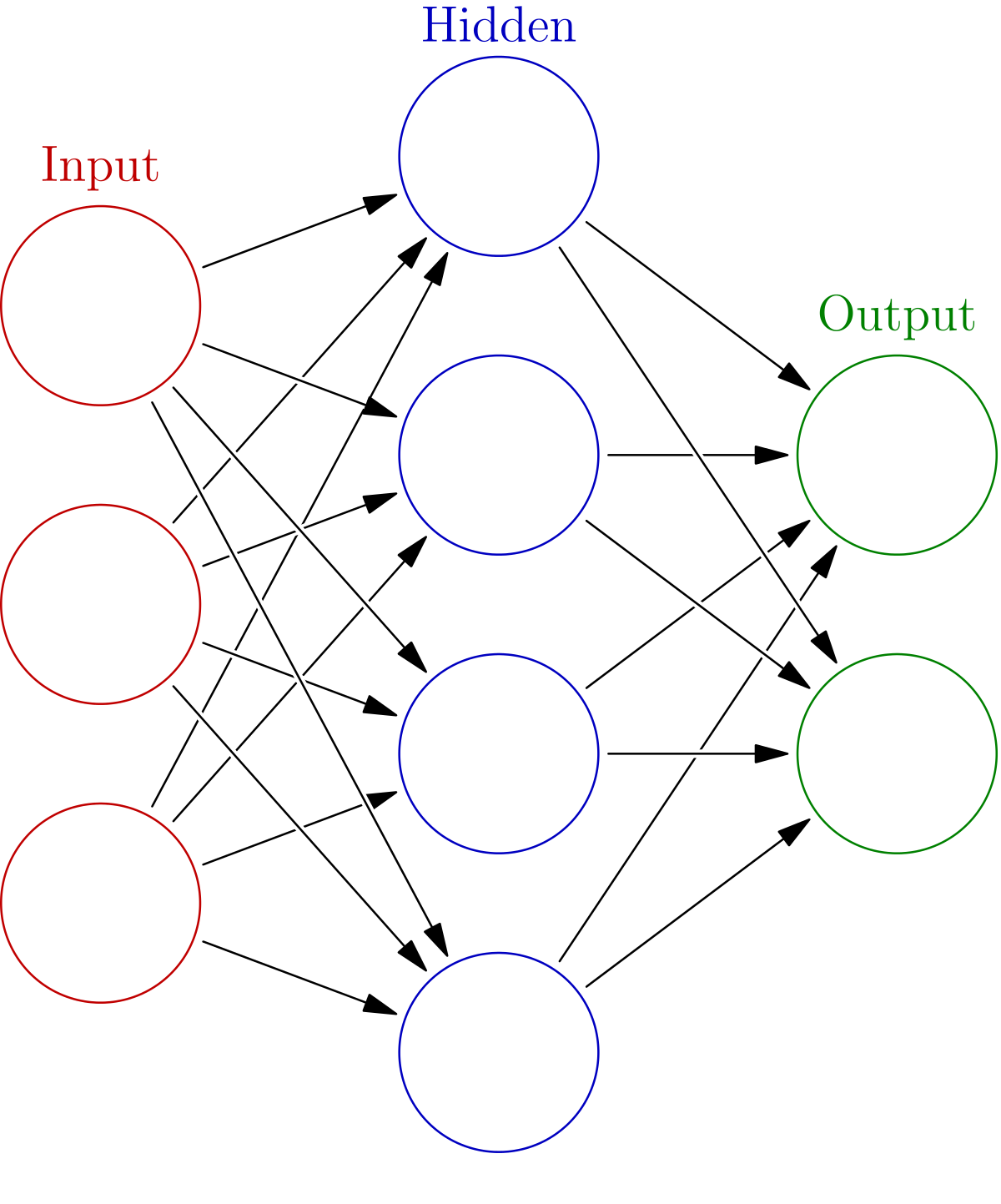}
\caption{
Schematic view of a simple feedforward ANN with one hidden layer. Circles represent neurons where information is processed and arrows represent travel direction of information through the network. 
}
\label{ann}
\end{center}	
\end{figure} 

\begin{figure*}
\begin{center}
\includegraphics[width=.9\textwidth, trim={0 0 5cm 0}]{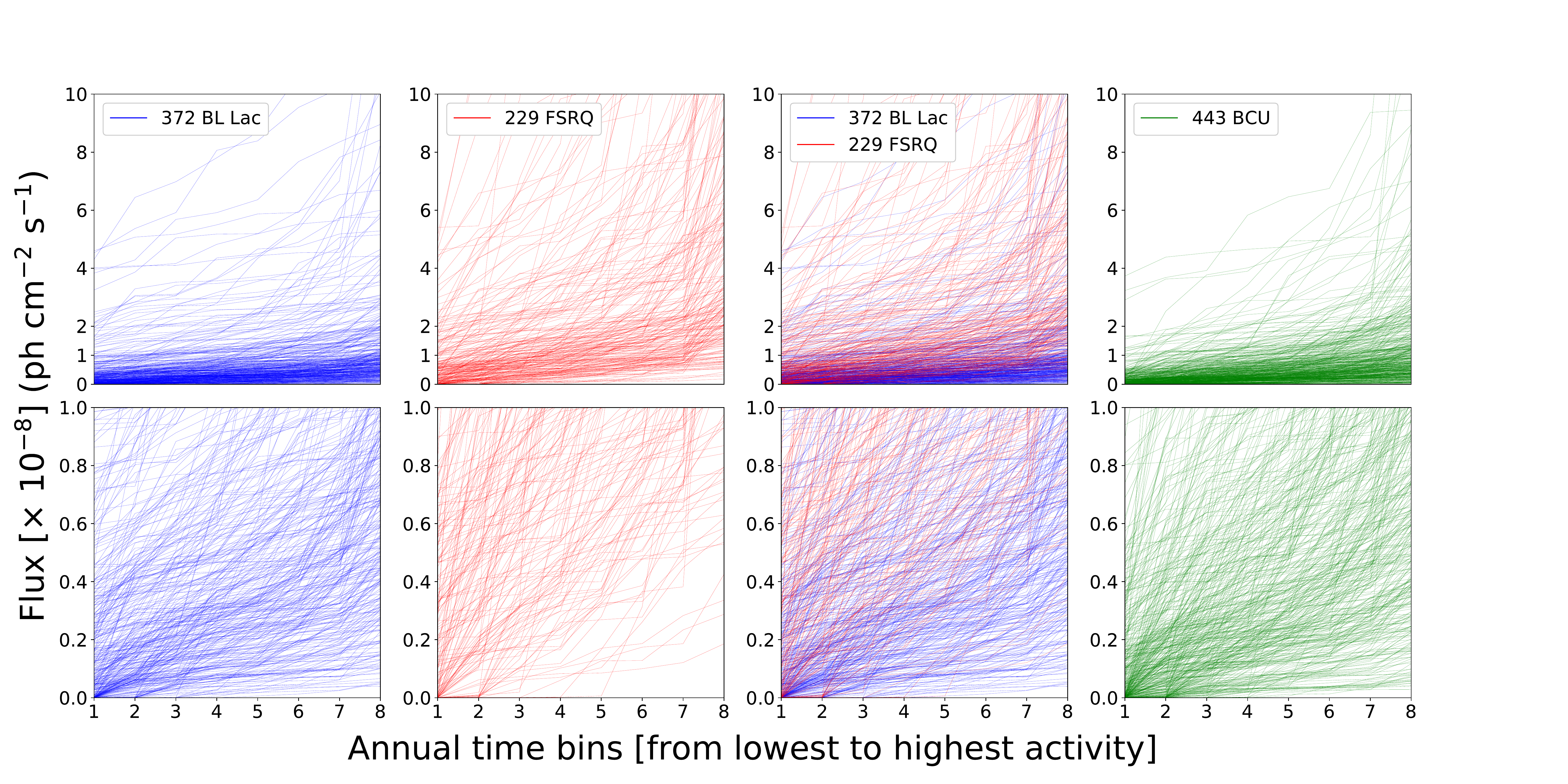}
\caption{
The annual fluxes of 4FGL blazars sorted from lowest to highest values. Each curve represents a single source. Vertical axes present annual flux values for the energy range 0.1--100 GeV. The lower and upper plots correspond to flux ranges of 0--1 $\times$ 10$^{- 8}$ ph cm$^{-2}$ s$^{-1}$ and 0--10 $\times$ 10$^{- 8}$ ph cm$^{-2}$ s$^{-1}$. Horizontal axes present 8 annual time bins. For each source the curve is made by sorting annual flux values from lowest (1st time bin) to highest (8th time bin). Therefore, lower time bins correspond to years of lower activity while higher to years of higher activity for each source. BL Lacs are in the first plot column (left-hand panel), FSRQs in the second, both are in the third and BCUs are in the fourth (right-hand panel). For clarity only one third of sources for each class are plotted. 
}
\label{FluxHistory}
\end{center}
\end{figure*}

The ANN technique is modeled by the way biological neural systems in the brain work. The schematic view of a simple ANN is presented in Fig.~\ref{ann}. The information enters the input layer and is sent to neurons in hidden layer(s) where it is processed. Finally it exits the output layer producing a desired outcome (classification of objects, for example).

Basically, ANN is a mathematical function over an $N$-dimensional space, where $N$ is the number of input parameters to the network. Input parameters are values which describe an object (blazars in our case). ANN produces a likelihood for the object to belong to a certain class (when ANN is used for classification). The network is trained on already classified objects (known BL Lacs and FSRQs in our case). Training the network involves adjusting the very large number of ANN parameters in order to find a function which best separates objects belonging to different classes. The network is then tested on classified objects which were not used in training in order to evaluate the trained network. After that the trained network can be used to classify unknown objects (BCUs in our case).

More detailed information on general characteristics of ANN, and particularly ANN for classifying BCUs, is present in \textit{C16}, \textit{S17}, \textit{K19}. The following method mostly follows the ones from the 3 cited works (particularly \textit{K19}).
Spectra and variability (obtained from the light curve) are two main features by which BL Lacs and FSRQs are distinguished in $\gamma$-ray band \citep{4fgl, 4lac}. Therefore, for input parameters we used $\gamma$-ray light curves and spectra present in the 4FGL catalog. More precisely we used 8 energy-integrated fluxes corresponding to 1-year observation periods sorted by increasing value, and time-integrated flux values in 7 different energy bands. This produced a set of $N=15$ input parameters to the network for each source.

\subsection{Gamma-ray light curves}
\label{<variability>}

We use the $\gamma$-ray light curves with sorted flux values from lowest to highest for each source, which is in line with an Empirical cumulative distribution function. In the 3FGL catalog, time bins had a duration of one month. This created a set of (12 months $\times$ 4 yr) 48 sorted monthly flux values for each source, which were used in previous studies. The 4FGL catalog contains light curves with a bin duration of 1 yr. This created a set of (1 yr $\times$ 8 yr) 8 sorted annual flux values for each source. While the light curves in the 4FGL catalog have smaller time resolution, each flux value is obtained from a 12 times longer observational period; therefore they are more precisely determined. Consequently, there are no undetermined fluxes with only upper limits in the 4FGL light curves as was the case with the 3FGL light curves. Also, the twice as long observational period allows us to better capture true characteristics of blazar light curves. Although the 4FGL also has two-month-long light curves, we choose to focus on the longer duration time bins for the reasons described above.

Sorting the flux values from lowest to highest is one way of making blazar activities comparable. The 8 annual time bins corresponding to 8 years of \textit{Fermi}-LAT observations are random time intervals in the life of each blazar. Fluxes in the same observational time bin go into the same input node of the network, but there is no physical meaning for this. By sorting the flux values, we are directly comparing fluxes of dimmest, average, and brightest periods for each blazar and relationships between them. 

The corresponding curves are presented in Fig.~\ref{FluxHistory}. Most of the sources occupy the range of flux values in the 0--10 $\times$ 10$^{-8}$ ph cm$^{-2}$ s$^{-1}$ interval (upper plots). In order to capture characteristics at lower flux values, the range 0--1 $\times$ 10$^{-8}$ ph cm$^{-2}$ s$^{-1}$ has been plotted separately below.

The curves contain information on average brightness, maximum annual-averaged activity, variability of sources, flaring patterns, etc. 
BL Lacs are on average dimmer than FSRQs in the \textit{Fermi}-LAT energy range. Their activity tends to be more continuous over time than that of FSRQs. Quick comparison between BL Lacs and FSRQs shows several features. In the lower right part of the plots there is an area where mostly BL Lacs are found. Sources passing through this area are ones which have lower flux ($\lesssim$ 1 $\times$ 10$^{- 8}$ ph cm$^{-2}$ s$^{-1}$) during their brightest years. Both dimmer and brighter BL Lacs, on average, have more \textit{horizontal} curves with respect to FSRQs (of similar average flux), which reflects their more continuous emission over time and lower variability.

Similar behaviour was present with 3FGL blazars with a few differences. In general the resolution is higher (time bins smaller) for 3FGL blazars, so the differences between BL Lacs and FSRQs are more obvious. For example, the area of lower flux values during brightest periods where mostly BL Lacs can be found is more clear for 3FGL BL Lacs ($\lesssim$ 2 $\times$ 10$^{- 8}$ ph cm$^{-2}$ s$^{-1}$) than for 4FGL BL Lacs. 3FGL BL Lacs and especially BCUs have large numbers of time bins, during dimmer periods, with only upper limits while 4FGL BL Lacs and BCUs have relatively small but defined flux values thanks to the larger time bins of 4FGL blazars. 

\begin{figure*}
\begin{center}
\includegraphics[width=.9\textwidth]{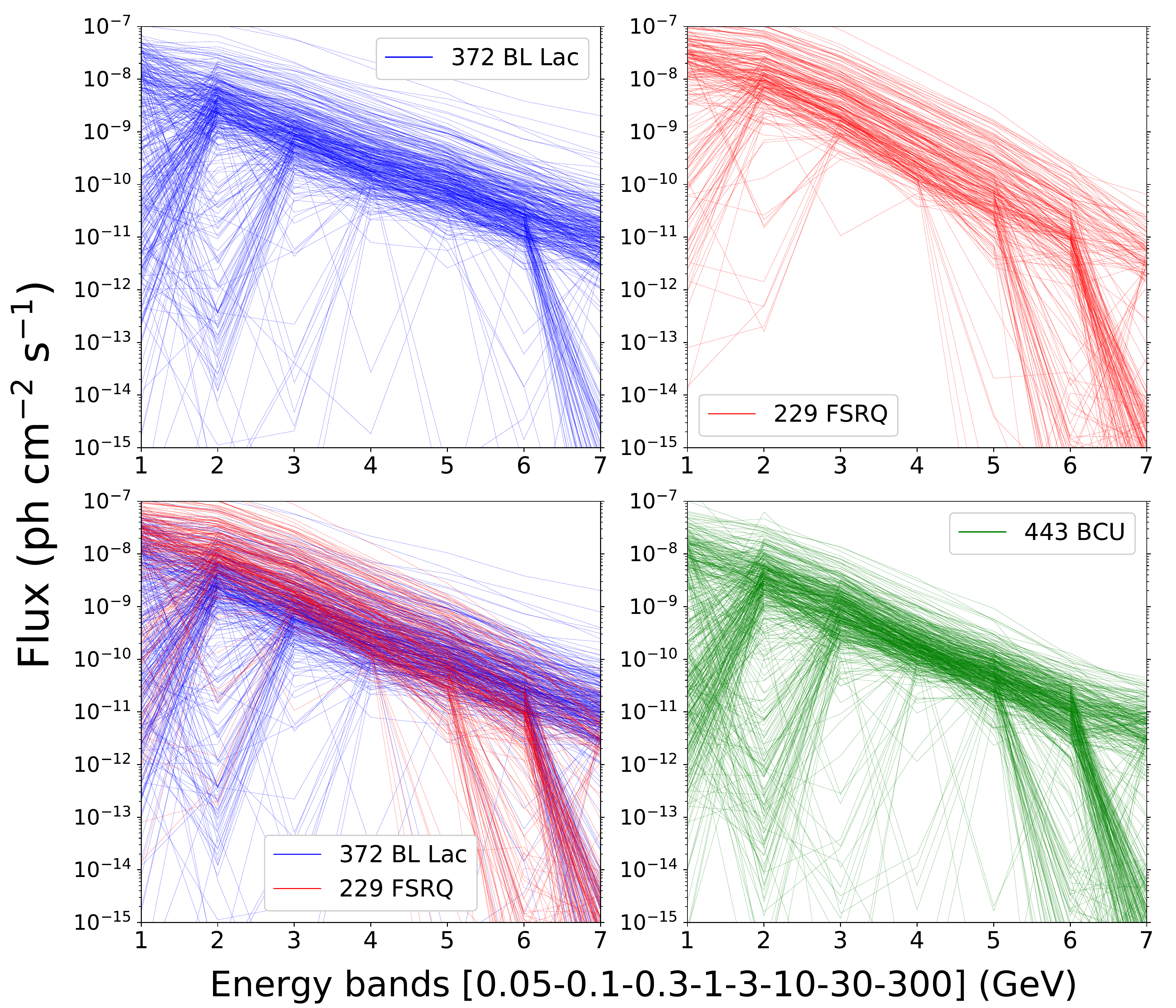}
\caption{
Time-integrated fluxes in 7 energy bands: Band 1: 0.05--0.1 GeV; Band 2: 0.1--0.3 GeV; Band 3: 0.3--1 GeV; Band 4: 1--3 GeV; Band 5: 3--10 GeV; Band 6: 10--30 GeV; Band 7: 30--300 GeV. Each curve represents a single source. BL Lacs (blue) are in the top-left, FSRQs (red) in the top-right, both are in the lower left and BCUs (green) are in the lower right. For clarity only one third of sources for each class are plotted.
}
\label{energybands}
\end{center}
\end{figure*}

\subsection{Gamma-ray spectra}
\label{<spectrum>}

We used spectral information in addition to light curves with sorted flux values. The 4FGL catalog contains time-integrated fluxes in 7 energy bands: 0.05--0.1, 0.1--0.3, 0.3--1, 1--3, 3--10, 10--30, 30--300 GeV (Fig.~\ref{energybands}). This is a wider energy range (0.05--300 GeV) than the one from the 3FGL catalog (0.1--100 GeV), which contained 5 energy bands. Energy bins 2, 3, 4, 5 (0.1--0.3--1--3--10 GeV) for 4FGL blazars are the same as energy bins 1, 2, 3, 4 for the 3FGL ones. Energy bin 1 (0.05--0.1 GeV) covers a new energy range in 4FGL while bins 6 and 7 (10--30--300 GeV) correspond partly to bin 5 in 3FGL (10--100 GeV). The improvement is due to longer observation period, i.e. better statistics and improvements in analysis techniques \citep{4fgl}. This set of parameters contains information of average spectral index, spectral curvature, spectral breaks, hardness ratios and other spectral information.

In the previous case fluxes were sorted in ascending order so that, among other reasons, there would be physical meaning for comparing fluxes (and relationships between them) that go into the same network input node. Here the fluxes of blazars in the same energy band go into the same network input node so the physical meaning is already there.

Quick comparison between BL Lacs and FSRQs shows several features: there is a difference in slope, i.e. average power-law index, with BL Lacs having a lower one; BL Lacs on average have higher flux values than FSRQs in the highest energy band and vice versa for lowest; some blazars show sharp breaks in slopes at lower and/or higher energy bands, and this behavior is mostly different for BL Lacs and FSRQs.

Comparing the spectral relationship of 4FGL BL Lacs to FSRQs with their relationship in 3FGL, it is mostly similar with several differences mainly related to spectral breaks thanks to the widening of the energy range. For example bin 1 in 4FGL (0.05--0.1 GeV) covers a new energy range and shows that some blazars peak in the energy range 0.1--0.3 GeV, which was not clear before. These blazars seem to be BL Lacs and FSRQs in similar proportion as the rest of the two classes. It also shows that some blazars (mainly BL Lacs) have a sharp decrease in flux from bin 1 to bin 2, and then sharp increase in bin 3, with the second feature also being present in 3FGL blazars.

\subsection{The Network}
\label{<network>}

Here we briefly describe the network architecture and the training strategy. They mostly follow the architecture and training strategy in \textit{K19} and are explained in more detail there, particularly how overfitting was handled.

We used 8 annual fluxes sorted in ascending order and 7 flux values in different energy bands as input parameters. This produces a $N=15$ dimensional parameter space in which each blazar occupies a certain position. We noted some obvious differences between BL Lacs and FSRQs when comparing their annual fluxes (Section~\ref{<variability>}) and spectra (Section~\ref{<spectrum>}). The purpose of the ANN algorithm is to fully determine the differences and to quantify them. It does so not just for sorted light curves and spectra separately but also taking into account relationships between them by examining the whole 15D parameter space.

The number of input neurons was 15 (8 for 8 annual sorted fluxes plus 7 for 7 fluxes in energy bands). The hidden layer had 40 neurons. The output layer had 2 neurons. The two output neurons produce likelihood that a source is BL Lac $L_B$ or an FSRQ $L_F$ such that $L_B + L_F = 1$ for each source. The larger the $L_B$, more likely that the source is a BL Lac and vice-versa. The Loss/Cost function used was the mean squared error. The number of ANN parameters, which are adjusted during network training, for this architecture is on the order of $\sim$700.

The training set consisted of 70\% and the test set of 30\% of the 4FGL classified blazars. The process of training the network and results from testing the network may depend on which sources were selected for the training sample and which for the testing sample. For this reason, we performed  training and testing the network on 300 different combinations of training and testing samples and compared the results.

\section{Network outputs}
\label{<3>}

\subsection {Test sample sources vs BCUs}
\label{<3.1>}

\begin{figure}
\begin{center}
\includegraphics[width=.45\textwidth]{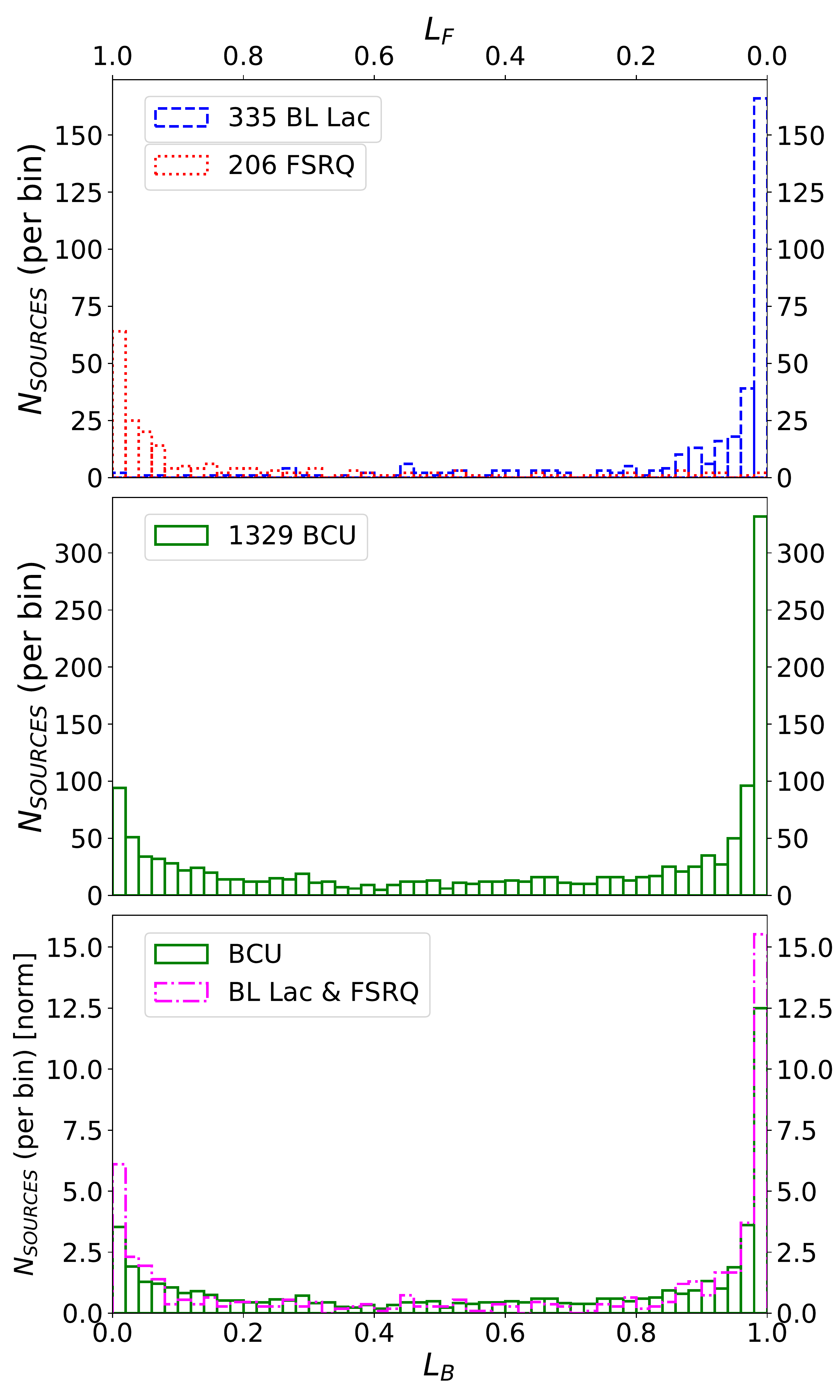}
\caption{ 
\textbf{Top:} Histogram of $L_B$ for BL Lac (blue) and FSRQ (red) sources from the test sample obtained from inputting test sample source parameters into the trained network. 
\textbf{Middle:} Histogram of $L_B$ for BCUs obtained from inputting BCU parameters into the trained network.
\textbf{Bottom:} Histogram of BCUs (green) and sum of BL Lacs and FSRQs from the test sample (purple). Both histograms are normalized such that surface of each equals 1 (the number of sources in both is the same).
} 
\label{<hist>}
\end{center}
\end{figure}

In order to better present the results of the full analysis, we show here results from a single train and test sample which is representative of the full analysis. The histogram of $L_B$ for BL Lacs and FSRQs from the test sample is shown in the upper plot in Fig.~\ref{<hist>}. It is obtained by inputting parameters of sources from test sample (which the network never "saw") into the trained network. As expected, BL Lacs concentrate towards $L_B \rightarrow 1$ while FSRQs $L_B \rightarrow 0$. The number of BL Lacs and FSRQs is 30\% of the total sample and the ratio of BL Lacs to FSRQs is the same as the ratio in the total sample.

\begin{figure}
\begin{center}
\includegraphics[width=.45\textwidth]{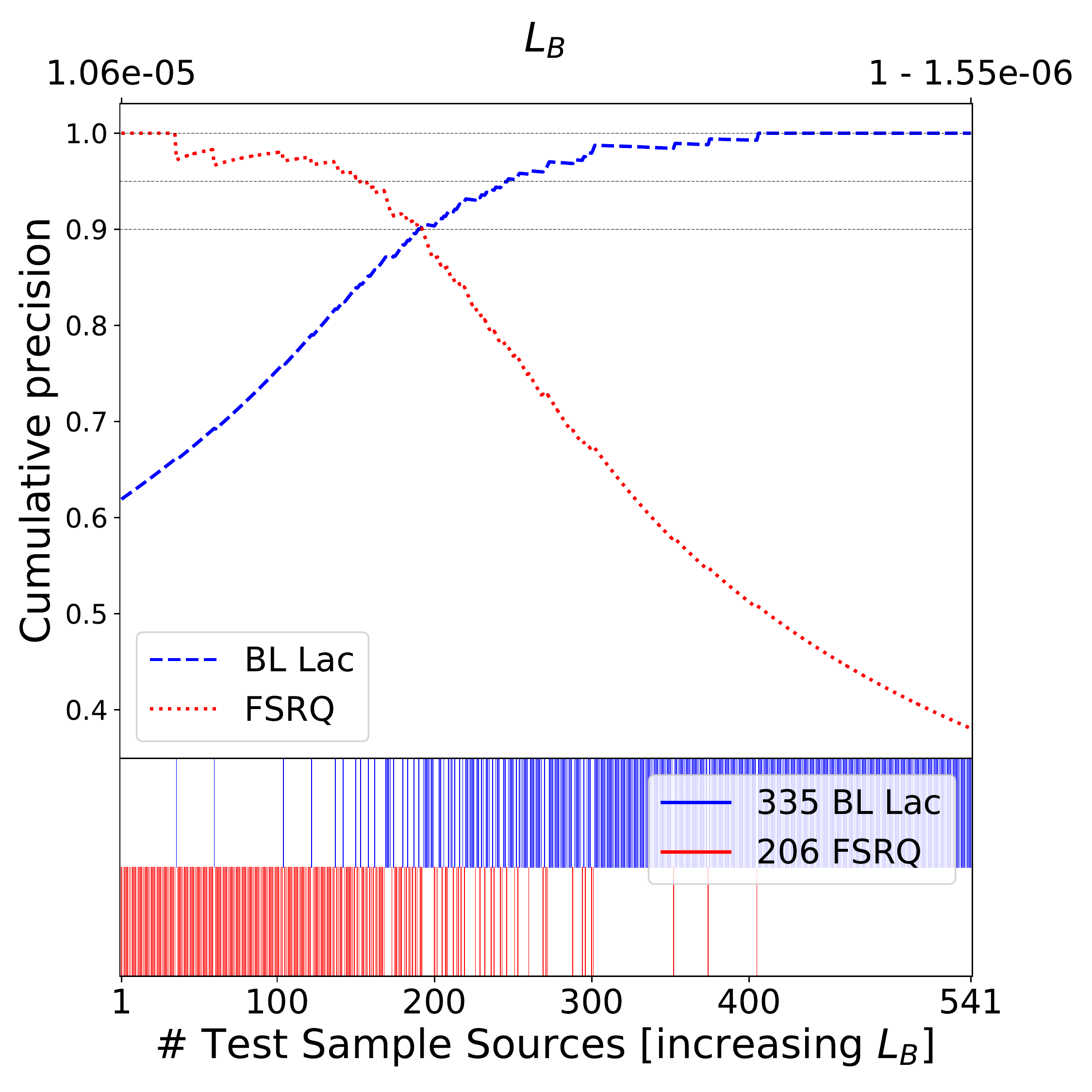}
\caption{
\textbf{Lower bar:} 335 BL Lacs (blue; vertical lines in upper half of the bar) and 206 FSRQs (red; vertical lines in lower half of the bar) from the test sample sorted by increasing $L_B$ and at equal horizontal distance from each other. The $L_B$ does not increase linearly in the plot. The lowest (left) and highest (right) obtained $L_B$ are shown in the upper plot. 
\textbf{Upper plot:} change of cumulative precision with the $L_B$ for BL Lacs (blue) and FSRQs (red).
} 
\label{<TestPrec>}
\end{center}
\end{figure}

In Fig.~\ref{<TestPrec>} the cumulative precision versus $L_B$ is shown. Sources from the test sample are sorted by their $L_B$.
The two curves practically meet at 90\% precision value, meaning that almost all sources from the test sample can be separated with 90\% precision.

Inputting BCUs into the trained network produces a histogram (middle plot in Fig.~\ref{<hist>}) with peaks towards $L_B \rightarrow 1$ and $L_B \rightarrow 0$, imitating the distribution of BL Lacs and FSRQs from the test sample (upper plot in Fig.~\ref{<hist>}). This is expected since the large majority of BCUs are either BL Lacs or FSRQs. We can expect that BCUs with large $L_B$ are mostly BL Lacs and vice versa.

In order to construct the same precision vs. $L_B$ relation (Fig.~\ref{<TestPrec>}) to BCUs, the BCU distribution with respect to $L_B$ should be as similar as possible to the combined distribution of BL Lacs and FSRQs from the test sample with respect to $L_B$. This is not entirely the case. In the bottom plot in Fig.~\ref{<hist>}, the histograms of BCUs and test sample sources are compared. Both histograms are normalized to the number of sources. While the peak at $L_B \rightarrow 1$ on histogram from the test sample sources and BCUs is very similar, the peak at $L_B \rightarrow 0$ is less pronounced for BCUs. In the middle range $0.2 \gtrsim L_B \gtrsim 0.8$ there are more sources with respect to both peaks for BCUs than for the test sample sources. In order to quantify these differences we use differential precision.

\begin{figure}
\begin{center}
\includegraphics[width=.45\textwidth]{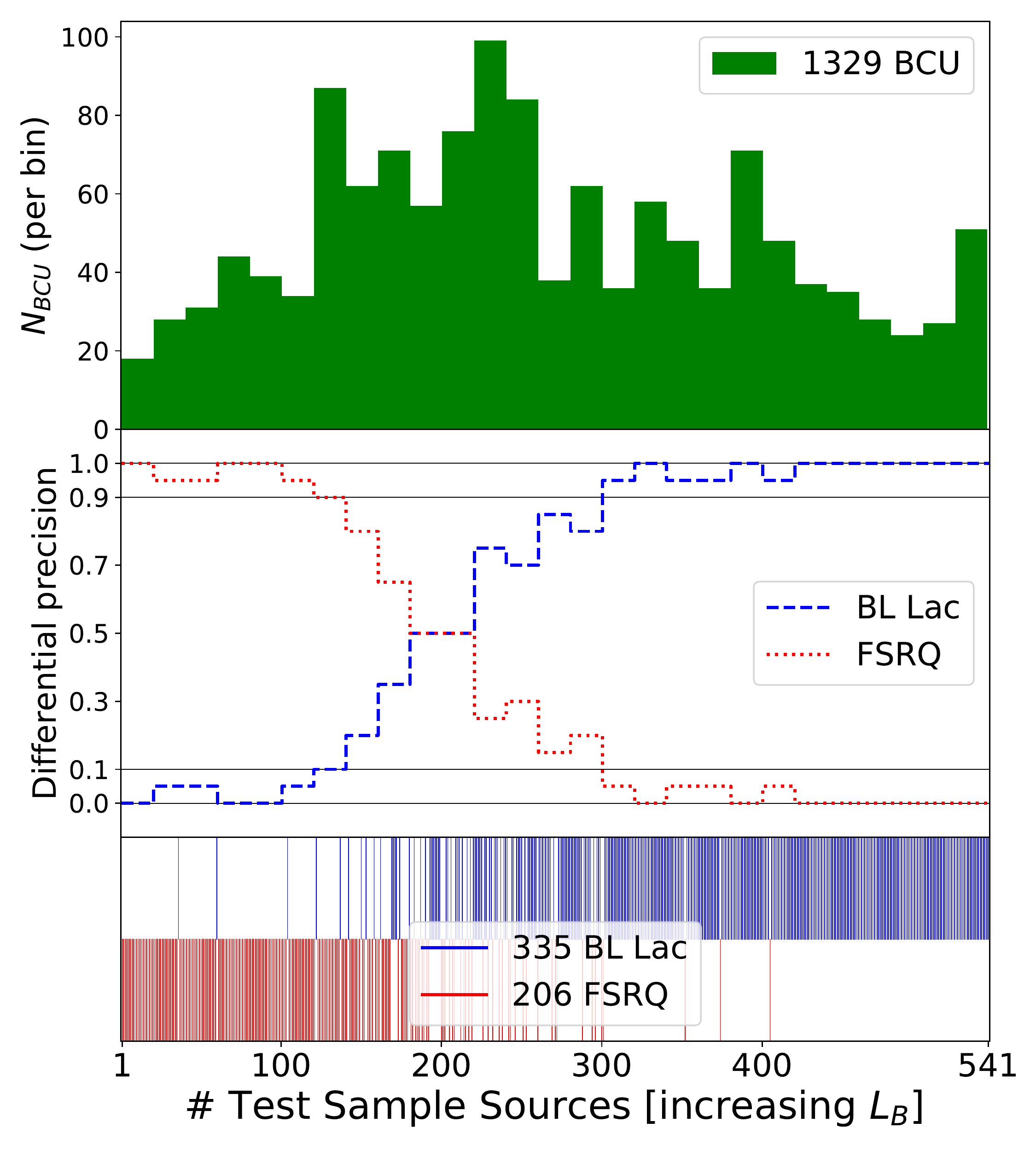}
\caption{
\textbf{Lower bar:} 335 BL Lacs (blue; vertical lines in upper half of the bar) and 206 FSRQs (red; vertical lines in lower half of the bar) from the test sample sorted by increasing $L_B$ and at equal horizontal distance from each other. The $L_B$ does not increase linearly.
\textbf{Middle:} differential precision for each bin which contains 20 test sample sources (last bin has 21). Differential precision is ratio of BL Lacs to all sources in each bin and vice-versa for FSRQs.  
\textbf{Top:} number of BCUs in each bin. Each BCU is assigned to a bin such that its $L_B$ is in between $L_B$ of the bin edges.
} 
\label{<TestPrecDiff>}
\end{center}
\end{figure}

In Fig.~\ref{<TestPrecDiff>} the differential precision, obtained from the test sample, is compared to $L_B$ of BCUs. The lower bar is the same as in Fig.~\ref{<TestPrec>} and presents test sample BL Lacs and FSRQs sorted by increasing $L_B$. The middle plot shows the differential precision $P_B$. It is obtained by binning sources from the lower bar in equal bins of 20 (the last bin has 21 sources). Then, $P_B$ is calculated for each bin as the ratio of BL Lacs to the number of sources and vice versa for FSRQs. This produces a set of $P_B$ (a step function) for BL Lacs (blue line) and FSRQs (red line) with resolution of 0.05 (1/20) such that their sum is 1 for each bin. Then the BCUs are sorted in each bin based on their $L_B$ (upper plot) and a value of $P_B$ (middle plot) is assigned to each BCU (upper plot). In this way $P_B$ of BCUs can be considered a as a probability that the given BCU is BL Lac or FSRQ.

Examining Fig.~\ref{<TestPrecDiff>}, the BCU distribution is not uniform across bins, meaning that the BCU distribution with respect to $L_B$ is not the same as that of the combined BL Lacs and FSRQs from the test sample, which number 20 in each bin. A larger than average number of BCUs are in the range where $P_B$ for either class is less than 90\%.

To overcome peculiarities of a single train and test sample and resolution lost to binning, the same process of training and testing the network was repeated for 300 different train-test samples. The final differential precision for each BCU $\bar{P_B}$ is then calculated as the average of 300 $P_B$ values. The value $\bar{P_B}$ can then be considered a probability of a given BCU to be BL Lac (or $\bar{P_F} = 1 - \bar{P_B}$ to be FSRQ) taking into account fluctuations due to train-test sample selections.

Lower and upper values of the error interval are 2 values corresponding to $\approx$16th and $\approx$84th percentile of 300 $P_B$ sorted from lowest to highest\footnote{Since all 300 values have resolution of 0.05, the same values were linearly extrapolated in [-0.025, +0.025] range.}. The interval in between these values can then be considered 1$\sigma$ errors due to differences in train-test sample selections.

We used differential precision to obtain probabilities for BCUs and classify them instead of thresholds obtained from test sample cumulative precision. Therefore we did not apply any cut to the test sample BL Lacs and FSRQs which occupy same parts of the parameter space (which makes them hardly distinguishable) in which BCUs are hardly present.

\subsection {Differential precision versus Likelihood}
\label{<3.2>}

\begin{figure}
\begin{center}
\includegraphics[width=.45\textwidth]{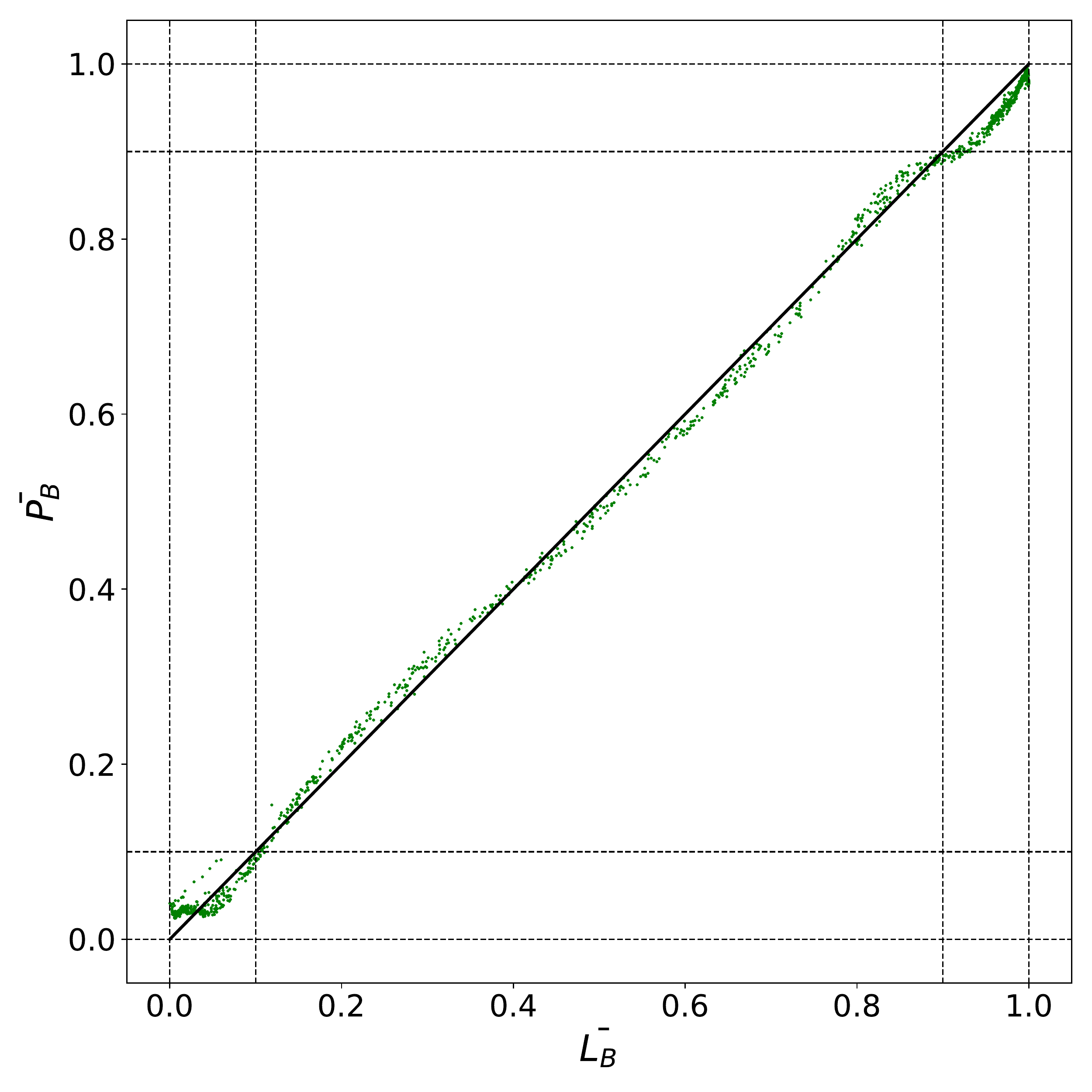}
\caption{
Average differential precision of 1329 BCUs to be BL Lac $\bar{P_B}$ with respect to average network output to be BL Lac $\bar{L_B}$.
}
\label{PvsL}
\end{center}
\end{figure}

When using mean squared error (MSE) \citep{gis90, ric91} or cross-entropy \citep{ric91} for Loss/Cost function, the network output $L_B$ can be considered an approximation to class probabilities. The accuracy of approximation depends on characteristics of the network architecture and training data \citep{ric91}. Since we used MSE, we also calculate $\bar{L_B}$ as an average of 300 $L_B$ and lower and upper limits as $\approx$16th and $\approx$84th percentile of 300 sorted $L_B$. The $\bar{L_F}$ value is just $\bar{L_F} = 1 - \bar{L_B}$. The comparison of $\bar{P_B}$ and $\bar{L_B}$ is shown in Fig.~\ref{PvsL}. Both quantities for all 1329 BCUs are very close in value. There are obviously some small systematic differences, but they do not change the overall results by much. The differences are probably due to resolution lost to binning for $\bar{P_B}$ and the above-mentioned approximation accuracy for $\bar{L_B}$. When experimenting with Loss/Cost functions which are not MSE or cross-entropy, the separation of test sample sources (Fig~\ref{<TestPrec>}) and $\bar{P_B}$ of BCUs remain similar while $\bar{L_B}$ of test sample sources and BCUs may change significantly. In these cases $\bar{L_B}$ cannot be considered in absolute terms as direct probability; instead it can be used in relative terms to compare sources to each other.

Showing equivalence between $\bar{P_B}$ and $\bar{L_B}$ in this case, from here on out we will use $\bar{L_B}$, since its interpretation as direct network output is more obvious and it gives more precisely defined errors (which are additionally affected by binning for $\bar{P_B}$).

\subsection {Caveats}
\label{<3.3>}

Here we note some caveats in the supervised learning approach. The issues have to do with how the parameters of known sources (known BL Lacs and FSRQs) compare to parameters of unknown (BL Lacs and FSRQs among BCUs), and this is related to astronomical observations.

As a simple example, known LAT BL Lacs are 44\% more present in the northern Galactic hemisphere than in the southern one because larger and better optical spectroscopic data, required to identify BL Lacs so that LAT blazars can be associated to them, are more available for the Northern hemisphere \citep{4lac}. The \textit{Fermi}-LAT sweeps the whole $\gamma$-ray sky continually, and there is no reason to think that the fraction of LAT BL Lacs is larger for the Northern hemisphere. If the Galactic latitude was used as a parameter, the machine learning algorithms would wrongly assume that BCUs in the Northern hemisphere are more likely to be BL Lacs.

Regarding the parameters used in this work, one of the obvious differences is that BCUs have lower flux values compared to known BL Lacs and FSRQs. This means that BCU population density in the parameter space is different than that of combined known BL Lacs and FSRQs. However this is not an issue since the ANN function is defined for each part of the parameter space. It just means that $\bar{L_B}$ of BCUs will be differently distributed than those of combined known BL Lacs and FSRQs, but they will still be accurate. What is important is that the fraction of unknown BL Lacs and FSRQs among BCUs is similar to the fraction of known BL Lacs and FSRQs in each part of the parameter space, and that is a potential caveat.

Another important factor is the redshift/distance. The parameters in the 4FGL catalog are observational parameters. A different redshift for the same source would change its flux values, observational time bin intervals, and energy bin intervals. It would, of course, be more accurate to take into account these effects in the analysis, but the majority of BCUs do not have measured redshift. In any case the difference in observational parameters does exist for BL Lacs and FSRQs. What is important is that unknown BL Lacs and FSRQs among BCUs have a similar redshift distribution as known BL Lacs and FSRQs. This is part of the previous requirement that the fraction of unknown BL Lacs and FSRQs among BCUs is similar to that of known BL Lacs and FSRQs throughout parameter space.

\section {Results}
\label{<4>}

In Table~\ref{<tab2>} an example of 7 classified BCU sources is shown. The complete list of 1329 BCUs is available in electronic format. The table contains Galactic coordinates, $\bar{L_B}$ and upper and lower values of the error interval.\\

\begin{table*}
\begin{center}
\caption{
Example of 7 classified BCU sources. The full list is available in electronic format. Columns: 4FGL name, Galactic latitude, Galactic longitude, $\bar{L_B}$, lower value of error interval $\bar{L_B}^{low}$, upper value of error interval $\bar{L_B}^{up}$.
}
\label{<tab2>}
\begin{tabular}{l|cc|ccc}
\hline																	
Name & $b$ (deg) & $l$ (deg) & $\bar{L_B}$ & $\bar{L_B}^{low}$ & $\bar{L_B}^{up}$ \\
\hline
4FGL J1224.7$-$8313 & -20.397 & 302.096 & 0.039 & 0.031 & 0.048 \\
4FGL J0804.5+0414 & 18.180 & 217.568 & 0.105 & 0.089 & 0.124 \\
4FGL J0914.1$-$0202 & 30.177 & 233.058 & 0.431 & 0.327 & 0.540 \\
4FGL J0709.0+4304 & 21.177 & 174.289 & 0.830 & 0.769 & 0.893 \\
4FGL J1514.6$-$2044 & 30.895 & 342.539 & 0.920 & 0.898 & 0.942 \\
4FGL J0538.2$-$3910 & -30.297 & 244.438 & 0.977 & 0.969 & 0.986 \\
4FGL J2251.7$-$43208 & -63.607 & 14.738 & 0.997 & 0.995 & 0.999 \\ 
\hline
\end{tabular}
\end{center}
\end{table*}

\subsection {Classification}

\begin{figure}
\begin{center}
\includegraphics[width=.45\textwidth]{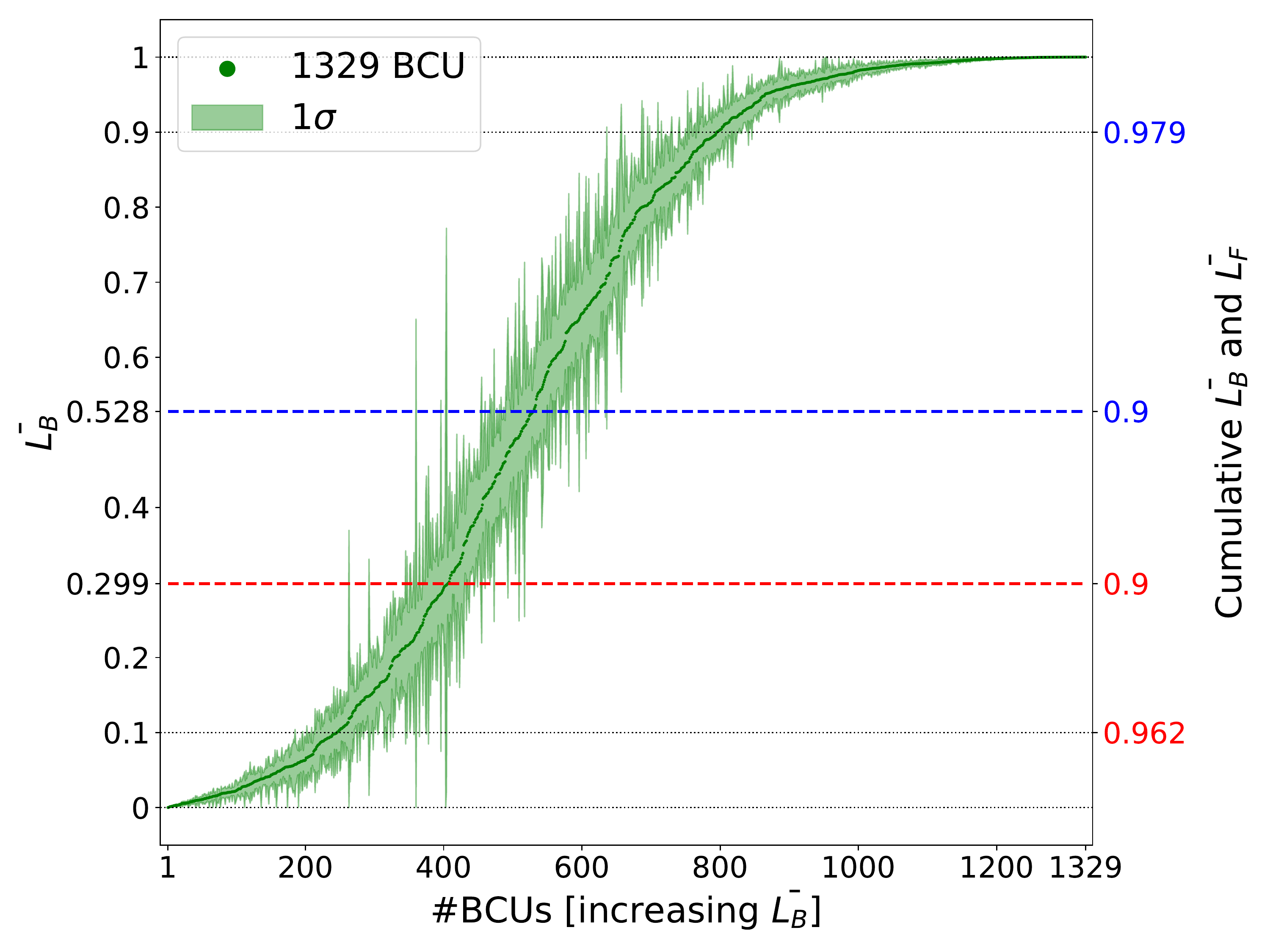}
\caption{
BL Lac probability $\bar{L_B}$ (1 - $\bar{L_F}$) of 1329 BCUs. Each BCU is presented by a green dot. The right-hand vertical axis shows the corresponding cumulative $\bar{L_B}$ (blue) and $\bar{L_F}$ (red). Values where cumulative $\bar{L_B}$ and $\bar{L_F}$ reaches 0.9 are marked by two horizontal blue and red dashed lines. Cumulative $\bar{L_B}$ and $\bar{L_F}$ values where $\bar{L_B} \geq 0.9$ and $\bar{L_F} \geq 0.9$ (high probable candidates) are also shown. The light green area corresponds to 1$\sigma$ error due to differences in train-test sample selections.
} 
\label{BCUPrecDiff}
\end{center}
\end{figure}

In Fig.~\ref{BCUPrecDiff} $\bar{L_B}$ of 1329 BCUs is shown along with the error. The quantities $\bar{L_B}$ and $\bar{L_F}$ (1 - $\bar{L_B}$) are probabilities for a BCU to be a BL Lac or a FSRQ. In order to present results as number of sources classified by precision metric, cumulative $\bar{L_B}$ ($\bar{L_B}_c$) and $\bar{L_F}$ ($\bar{L_F}_c$) are calculated as average $\bar{L_B}$ of all BCUs which have the same or higher $\bar{L_B}$ and vice-versa for cumulative $\bar{L_F}$. Cumulative values are shown on the right-hand vertical axis in Fig.~\ref{BCUPrecDiff}.

Selecting BL Lac and FSRQ candidates with a 90\% precision metric ($\bar{L_B}_c \geq 0.9$ and $\bar{L_F}_c \geq 0.9$; $\bar{L_B} \geq 0.528$ and $\bar{L_F} \geq 0.701$), 801 BCUs are classified as BL Lacs and 406 as FSRQs, leaving 122 unclassified. If only highly probable candidates are selected ($\bar{L_B} \geq 0.9$ and $\bar{L_F} \geq 0.9$; $\bar{L_B}_c \geq 0.979$ and $\bar{L_F}_c \geq 0.962$), then 534 BCUs are classified as BL Lacs and 245 as FSRQs. The second classification corresponds to 98\% precision for BL Lacs and 96\% for FSRQ.

The ratio of BL Lac to FSRQ candidates is about 2. For 90\% precision candidates it is 1.7, and for high probability candidates 2.2. Looking at Table~\ref{<fgl>}, it is clear that the ratio of known BL Lacs to FSRQs has steadily increased (1FGL: 1.1; 2FGL: 1.2; 3FGL: 1.4; 4FGL: 1.6). Since BL Lacs are on average dimmer in $\gamma$-rays than FSRQs, at first they were hard to detect but as \textit{Fermi}-LAT sensitivity increased due to its longer observational period, more BL Lacs started to be discovered with respect to FSRQs. For this reason, it is reasonable to assume that the true ratio among BCUs is larger than the current ratio of known BL Lacs to FSRQs.

Looking at Fig.~\ref{BCUPrecDiff}, the network can classify many more BCUs as almost certain BL Lacs ($\bar{L_B} \rightarrow 1$) than FSRQs ($\bar{L_F} \rightarrow 1$). This is because some BL Lacs occupy parts of parameter space where there are no FSRQs, i.e. certain group of BL Lacs are easily distinguishable from FSRQs.

The error is naturally small for sources with high $\bar{L_B}$ or $\bar{L_F}$ and the network classified them as probable BL Lac/FSRQ irrespective of train-test sample selection. For sources with intermediate $\bar{L_B}$, errors are larger. This is expected because classification of BCUs with properties (input parameters to the network) not clearly corresponding to either class will be more affected by fluctuation due to train-test sample selection.

\subsection {Classification vs galactic latitude}

The number of known BL Lacs and FSRQs within the Galactic plane region $|b| < 10^{\circ}$ is about 5\%. The number of BCUs within the $|b| < 10^{\circ}$ region is 18\%. The optical spectroscopy which is required to fully classify blazars is harder to do for sources near the Galactic plane.

\begin{figure}
\begin{center}
\includegraphics[width=.5\textwidth]{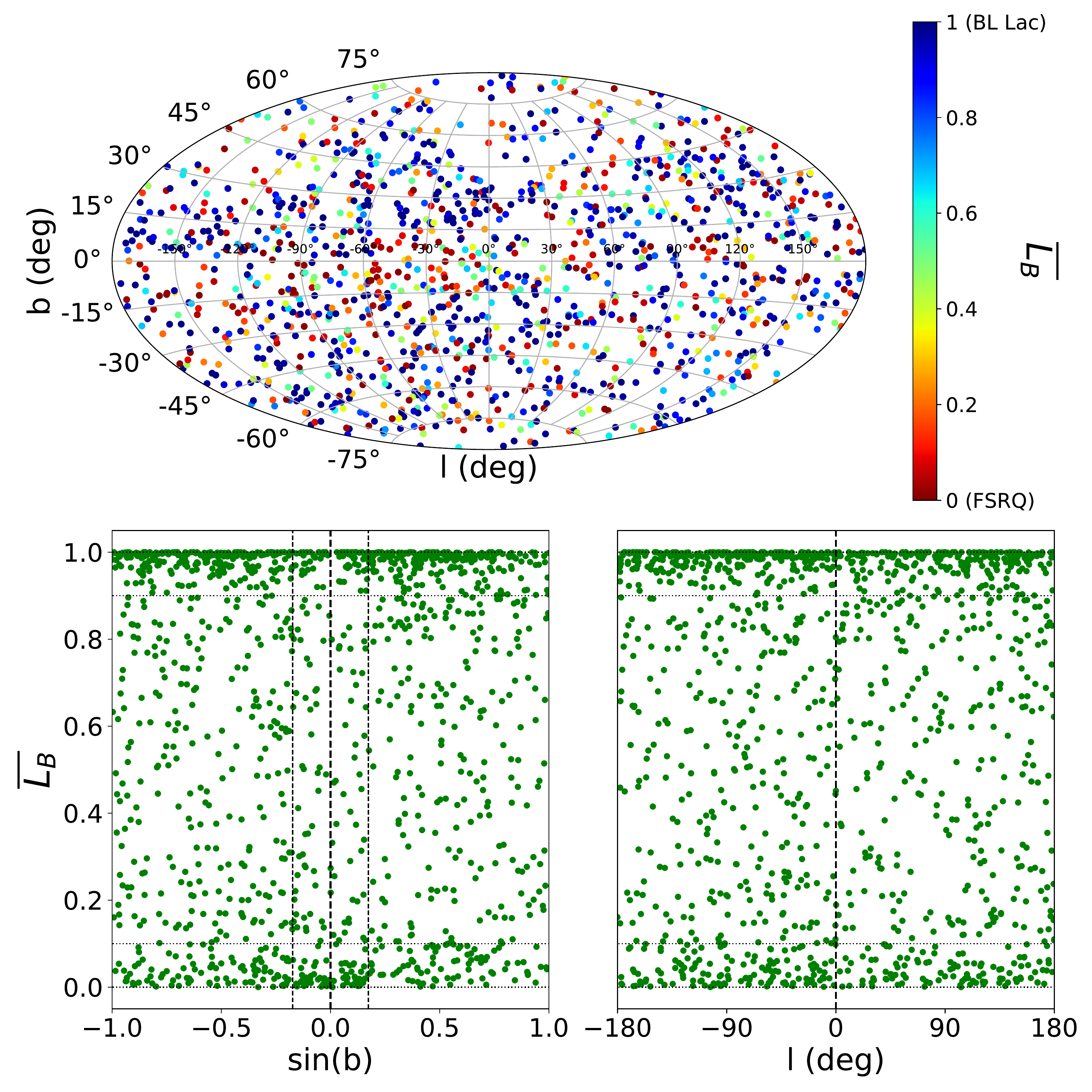}
\caption{
\textbf{Upper plot:} sky distribution in Galactic coordinates of 1329 BCUs from the 4FGL catalog. Colors correspond to $\bar{L_B}$. \textbf{Bottom plots:} $\bar{L_B}$ vs. Galactic longitude (left-hand panel) and latitude (right-hand panel). \textbf{Bottom-left plot:} the two black dashed vertical lines around $b = 0^{\circ}$ correspond to $|b| = 10^{\circ}$.
}
\label{<bcumap>}
\end{center}
\end{figure}

In Fig.~\ref{<bcumap>}, the sky distribution in Galactic coordinates of 1329 BCUs is shown together with their classification $\bar{L_B}$. Galactic diffuse $\gamma$-ray emission from the Galactic disk and many point and extended sources inside it make it more difficult to detect $\gamma$-ray blazars and measure their flux. Here we look at differences in classification between sources inside the Galactic diffuse emission area ($|b| < 10^{\circ}$) and those outside ($|b| > 10^{\circ}$).

The threshold of $\bar{L_B} \approx 0.42$ corresponds to a  precision of about 87\% at which all BCUs can be classified (865 BL Lacs and 464 FSRQs). Then MSE is defined as $\sum (1 - \bar{L_B})^2 / N$ for BL Lac candidates ($\bar{L_B} > 0.42$) and $\sum (0 - \bar{L_B})^2 / N$ for FSRQ candidates ($\bar{L_B} < 0.42$). This quantity is an average measure of uncertainty of BCUs classification, i.e. how far away $\bar{L_B}$ of BCUs is from the peaks at $\bar{L_B}$ = \{0,1\}. We found that this value is not bigger for BCUs at $|b| < 10^{\circ}$ than the ones at $|b| > 10^{\circ}$, meaning that BCUs near the Galactic plane are not classified with less certainty by the network. 

The average integrated (in time and energy) flux value of BL Lac and FSRQ candidates near the plane region is about two times larger than for candidates outside it. The same is true for known BL Lacs and FSRQs. This is expected since the $\gamma$-ray emission from the disk makes it harder to detect sources with lower flux. Known FSRQs have on average larger flux than known BL Lacs and the same is true for FSRQ and BL Lac candidates. For this reason, the fraction of FSRQ candidates inside the region ($112:464 \approx 0.24$) is larger than the fraction of BL Lacs ($125:865 \approx 0.14$). Considering only highly probable candidates ($\bar{L_B} > 0.9$ and $\bar{L_F} > 0.9$) the difference increases to 0.28:0.14. Therefore BCUs near the Galactic plane are made of a larger fraction of FSRQs when compared to BCUs outside of it. While the ratio of BL Lac to FSRQ candidates is about 2 for the whole sky, it is about 1 for the Galactic plane.

\section {Validation}
\label{<5>}

\begin{figure}
\includegraphics[width=0.45\textwidth]{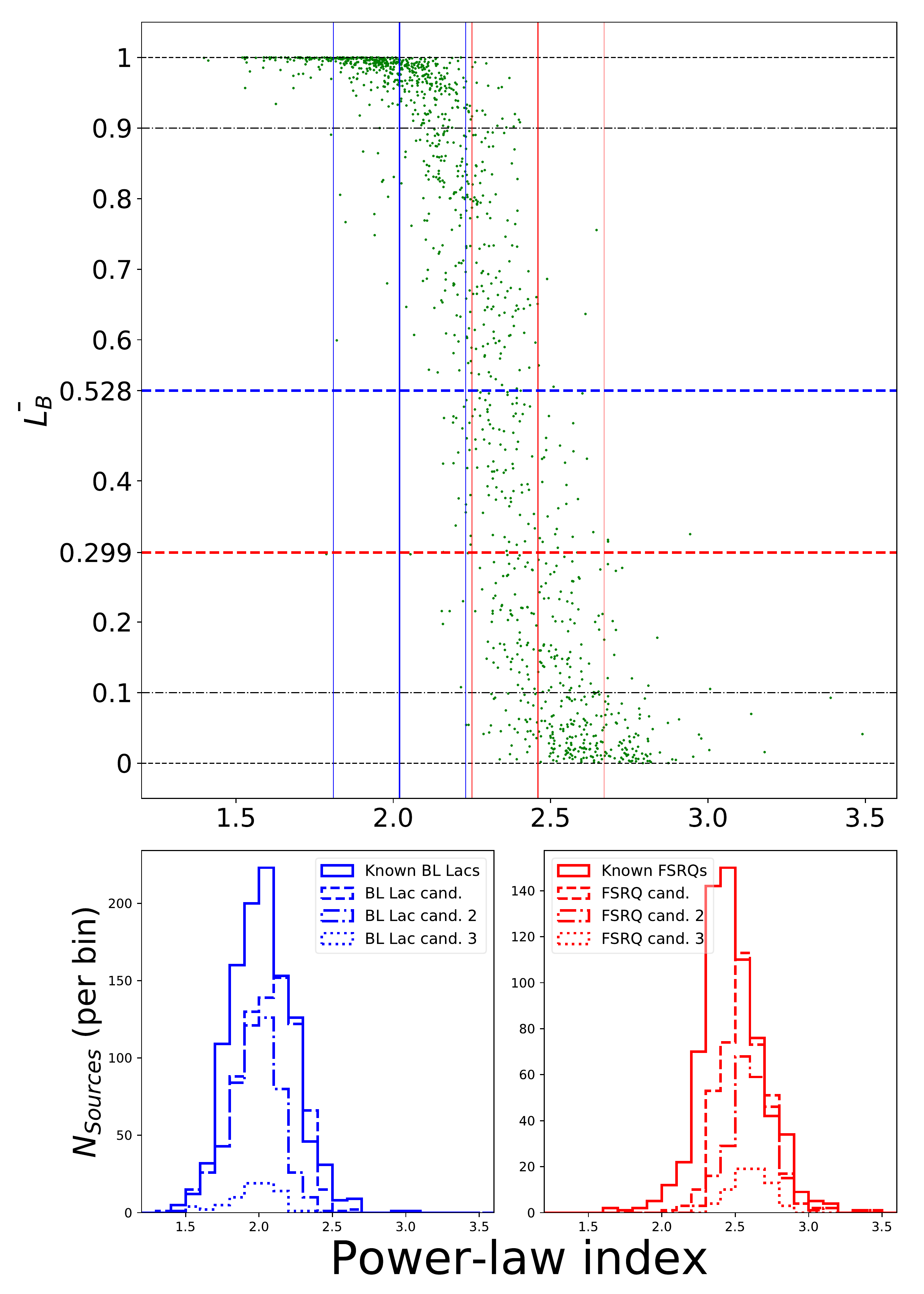}
\caption{
Comparison of BCU classifications with power-law (PL) indexes. The higher the probability of a BCU to be a BL Lac $\bar{L_B}$, the lower the PL index and vice-versa. The distribution of PL indexes of BL Lac and FSRQ candidates is in agreement with distribution of PL indexes of known BL Lacs and FSRQs. 
\textbf{Upper plot:} $\bar{L_B}$ of 1329 BCUs with respect to their PL indexes. The blue and red vertical lines are mean values of PL indexes of known BL Lacs (2.02) and FSRQs (2.47) and their 1$\sigma$ distribution widths which in both cases is 0.21. Horizontal blue and red dashed lines correspond to BL Lac and FSRQ candidates with 90\% precision metric.
\textbf{Bottom plots:} Histograms of PL indexes for BL Lacs (left-hand panel) and FSRQs (right-hand panel). BL Lac and FSRQ candidates are selected such that precision value is 90\%. Candidates number 2 are highly probable candidates ($\bar{L_B} > 0.9$ for BL Lacs and $\bar{L_F} > 0.9$ for FSRQs). Candidates number 3 are highly probable candidates which are confined to Galactic plane region $|b| < 10^\circ$.
}
\label{LvsPLI}
\end{figure}

It was discovered that BL Lacs and FSRQs are characterized by different $\gamma$-ray spectral properties. Usually BL Lacs show harder spectra than FSRQs \citep{3lac, 4lac}. Fitting 4FGL blazars, assuming a power-law (PL) spectral model, it was observed that the best-fit photon spectral index distribution is rather dissimilar for the two subclasses, making this observable an important $\gamma$-ray parameter to distinguish the two blazar classes. Since we did not include this parameter in our algorithm\footnote{We did use fluxes in different energy bands which contain information on average power-law index, but the power-law index per se was never used as an input parameter.}, in order to validate the performance of our algorithm (as a sanity check), we compared the PL index distribution of BCUs vs their $\bar{L_B}$ together with the PL distribution of known BL Lacs and FSRQs. 

A clear correlation between $\bar{L_B}$ and PL index of BCUs exists (upper plot in Fig.~\ref{LvsPLI}) such that higher $\bar{L_B}$ corresponds to lower PL index, i.e. harder spectrum, which is expected. Mean values and 1$\sigma$ spread for known BL Lacs and FSRQs is also shown. In the bottom plots in Fig.~\ref{LvsPLI}, the same correlation is shown in the form of histograms. BL Lac and FSRQ candidates follow the PL index distribution of known BL Lacs and FSRQs. High probability BL Lac candidates have even lower PL index and vice-versa for FSRQs, which is expected. Finally, high probability candidates within the Galactic plane region $|b| < 10^\circ$ follow the same distribution as high probability candidates in total, showing that correctness of classification is no different for the Galactic plane even though there are differences when it comes to integrated flux values of blazars. 

\begin{figure}
\includegraphics[width=0.45\textwidth]{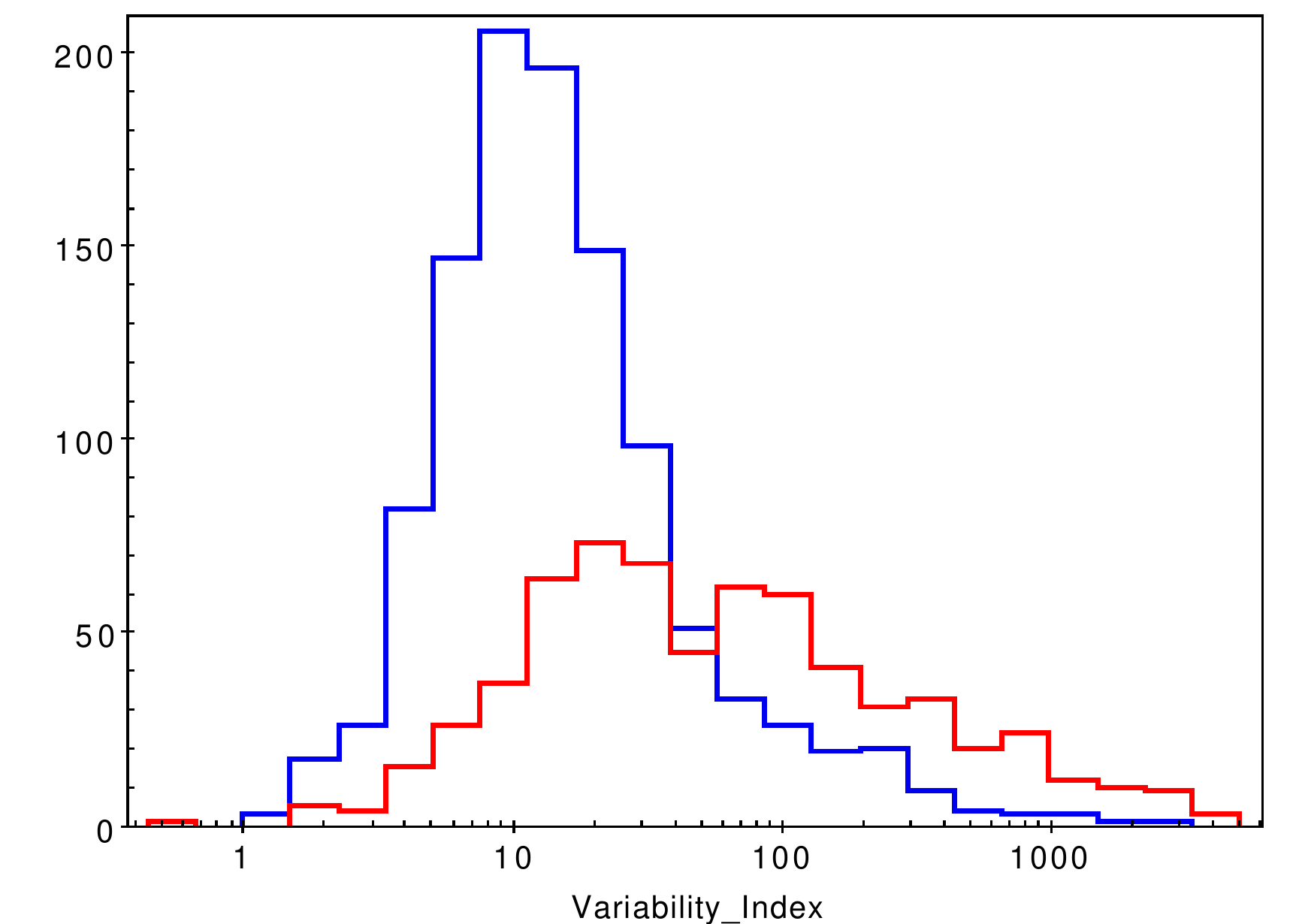}
\caption{
Variability index distribution for the known 4FGL blazars: BL Lacs (blue histogram) and FSRQs (red histogram). The evident overlap of the histograms show it to be inefficient at distinguishing blazar subclasses.
}
\label{piv}
\end{figure}

The good agreement of the PL index distribution for our candidates with the PL index distribution for known blazars confirms the correctness of the algorithm. We also plot the Variability Index distributions for known BL Lacs and FSRQs (Fig.~\ref{piv}). This parameter, unlike the PL index, is not as efficient at distinguishing blazar subclasses, so we did not use it for validation.

\section {Conclusion}
\label{<6>}

In this study we used a neural network method for the classification of uncertain blazars. We studied effects of selecting different training and testing samples, differences in test sample and BCU sample and discussed the meaning of network outputs. In the end, classification probabilities for each of 1329 BCUs are obtained along with error due to train-test sample selection. In terms of number of classified sources, 1207 BCU are classified compared to 1329 original BCUs, classifying 91\% of the sample with 90\% precision. Ratio of BL Lac candidates to FSRQ candidates is about 2:1 for the whole sky, and 1:1 for the Galactic plane. This result confirms that machine learning techniques are powerful methods to classify uncertain astrophysical objects and particularly blazars. 

In this work we used sets of $\gamma$-ray parameters that are spectra and light curves since these two features are known to be different for BL Lacs and FSRQs. It is, of course, possible to use other $\gamma$-ray parameters from the 4FGL catalog (including the PL index\footnote{Information on PL index is already contained in 7 fluxes in energy bands so it is not expected to bring new information. However the PL index is obtained from the likelihood fit over whole energy interval and it might be different for some blazars than PL index that would be obtained from 7 fluxes in energy bands.}) as well as multiwavelength data, such as X-ray and radio flux\footnote{It is possible to use radio and X-ray flux even if they are present for only a subset of LAT blazars (K19).} present in the upcoming Fourth Catalog of Active Galactic Nuclei \textit{4LAC} \citep{4lac} or other catalogs. This can be addressed in a future appendix to this paper.

Due to the increasing number of uncertain blazars during the \textit{Fermi}-LAT mission, the ANN technique could be a very worthwhile opportunity for the scientific community to quickly select promising targets for multiwavelength rigorous classification and related studies at different energy ranges, mainly at very high energies by the present generation of Cherenkov telescopes and the forthcoming Cherenkov Telescope Array\footnote{\url{www.cta-observatory.org.}} \citep{cta}.

\section {Acknowledgments}
            
The authors would like to thank David J. Thompson (NASA Goddard Space Flight Center, Greenbelt, MD, USA) for review of the paper. For science analysis during the operation phase we acknowledge the {\it Fermi}-LAT collaboration for making the results available in such a useful form. The authors also acknowledge their institutions for providing opportunity to carry out research. We thank the anonymous reviewer for suggestions leading to improvement of this work.

\label{lastpage}
\end{document}